\documentclass[twoside]{article}
\usepackage{fleqn,espcrc2}

\usepackage{graphicx}

\title{Is the Quantum Melting of a Polaron Wigner Crystal an
Insulator-to-Superconductor Transition ?}

\author{P. Qu\'emerais\address{Laboratoire d'Etudes de Propi\'et\'es
Electronique des Solides, CNRS, \\
        BP 166, 38042 Grenoble, France}
       and
        S. Fratini\address{INFM and Dipartimento di Fisica, 
        Universit\`a di Roma ``La Sapienza'' \\
        Piazzale Aldo Moro 2, 00185 Roma, Italy}}

\begin{document}

\begin{abstract}
On examining the stability of a Wigner Crystal (WC) in an ionic dielectric,
two competitive effects due to Polaron formation are found to be important: 
(i) the screening of the Coulomb forces which destabilizes the crystal,
compensated by (ii) the increase of the carrier mass (polaron mass).
The quantum melting of the Polaron Wigner Crystal (PWC) is examined.
By calculating the quantum fluctuations of both the electrons and the
polarization, we show
that there is a competition between the dissociation of the Polarons at
the insulator-to-metal transition (IMT), and a melting towards a polaron
liquid. We find that at strong coupling ($\alpha > \alpha^*$), a liquid
state of
polarons cannot exist, and the IMT is driven by polaron dissociation.
Next, we show that the dipolar interactions between localized polarons
are responsible for a phonon instability of the PWC as the density increases.
This provides a new mechanism for the IMT in doped dielectrics. Examining the
sign of the dielectric constant of the PWC, we conjecture that such an
instability could yield an Insulator-to-Superconductor transition.
\vspace{1pc}
\end{abstract}

\maketitle

\section{INTRODUCTION.}

The possibility of having a Wigner crystallization of large polarons in
the cuprates was first proposed in 1989 by Remova and Shapiro
\cite{remova}. This problem was independently studied by
us on purely theoretical grounds \cite{quem1,quem2} in the framework
of the Fr\"ohlich model, regardless of the
particular structure of the cuprates.
Up to now, the many-polaron problem was only studied in the high density
(metallic) limit \cite{mahan,lemmens}.
Recently, it was found \cite{defilipis} that a CDW instability can 
occur for sufficiently low densities.
Our approach relies on the opposite low density
limit, which we believe to be relevant for doped ionic dielectrics.

Let us consider a continuous polar medium characterized by
three parameters: the static dielectric constant $\varepsilon_s$, the
high frequency dielectric constant $\varepsilon_{\infty}$, and the
longitudinal optical phonon frequency $\omega_{LO}$. 
Reasonable estimates in the cuprates
(e.g. in La$_2$CuO$_4$ \cite{chen}) give $\varepsilon_{s} \approx 30$,
$\varepsilon_{\infty} \approx 5$ and $\omega_{LO} \approx 20-40 meV$.
Importantly, the static dielectric constant is almost isotropic in
all directions of space \cite{chen}. This experimental fact
means that the Coulomb forces and the phonons
carrying the polarization field act in the same way in all directions. On
doping such a polar material, the optical phonons dress and
screen the doping charges (this is the well-known polaron effect). The
strength of the electron-phonon coupling is given by the
dimensionless constant $\alpha =\left( {{{m^{*}}/{2\hbar ^{3}
\omega _{LO}}}}\right) ^{1/2}{{e^{2}}/{\tilde{\varepsilon}}}$, where
$1/\tilde{\varepsilon}=1/\varepsilon _{\infty}-1/\varepsilon _{s}$ and $m^*$ 
is the band mass. 
Due to their layered structure, the carrier masses in the cuprates are
highly anisotropic (in particular, no coherent motion takes place
along the \textit{c-}axis). In our model, 
this situation can be qualitatively described by introducing
a second parameter $m_c \gg m^*$, where $m^*$ now refers to the Cu-O
planes only.  One usually estimates $m^*
\approx 1-2 m_e$ ($m_e$ being the bare electron mass), so that  the
e-ph coupling in the \textit{ab-}planes is roughly $\alpha
\approx 4-5$.

As is well established, 
two polarons repel at distance $d$ as $1/ {\varepsilon_s} d$, provided
that the dielectric constants
satisfy $\varepsilon_{\infty}/\varepsilon_s >\approx 0.1$
\cite{vinetskii}. Therefore, at low density and neglecting 
the details of chemical doping, \textit{the crystallized state
is necessarily the ground-state of the many-polaron problem}.
This basic result has  several important consequences:

(i) At zero temperature, as the density increases, there is a
\textit{competition} between the quantum melting of the 
PWC towards a degenerate polaron liquid,
and the dissociation (ionization) of the polarons themselves.

(ii) At large values of the e-ph coupling $\alpha$, the melting towards a
polaron liquid is not possible, and it is \textit{the polaron
dissociation that drives the transition} \cite{quem1}.

(iii) Taking into account the dipolar interactions between polarons,
it can be shown that the PWC has an intrinsic \textit{phonon
instability} above a
given critical density $n_{c1}$, which is closely related to 
the polaron dissociation.

(iv) Such an instability has an optical signature: \textit{the peak due to the
existence of polarons in the optical conductivity, which is centered at a
definite frequency at very low density, is progressively shifted
towards lower values as the transition is approached, and eventually 
saturates at
$\omega_{LO}$} \cite{quem2}. This softening has  been
experimentally observed in Nd$_{2-x}$Ce$_x$CuO$_{4-y}$ \cite{lupi}.

(v) The longitudinal dielectric constant
$\varepsilon ({\bf k}, \omega)$ of the PWC is negative
on a large region of $({\bf k}, \omega)$ \cite{quem2}.  In particular,
$\varepsilon({\bf k},0)<0 $. 
As already noted in \cite{quem2}, the repulsive Coulomb interactions between
free charges liberated above the
critical doping would be 
\textit{overscreened} by the high frequency vibrations of
the localized PWC.

\section{POLARON DISSOCIATION: A NEW QUANTUM CRITICAL POINT.}

Let us focus on the quantum
melting (at zero temperature) of the PWC.
When the
e-ph coupling vanishes, we are left with an ordinary WC of electrons.  As
the density increases, the quantum fluctuations of the localized particles
increase and when they reach a certain magnitude, the crystal melts
towards a liquid state of electrons.  The effect of the host lattice
is negligible in that
case.  In the opposite limit $\omega_{LO} \rightarrow 0$, the
e-ph coupling $\alpha$  becomes infinite.  Each electron
in this limit is self-trapped in a polarization
potential well which is {\it frozen}.  As it is well-known, the latter has
a {\it coulombic} nature and behaves as $1/{\tilde \varepsilon }r$ at large
distance.  Since the polarization cannot move, the polarons, i.e.  the
electron {\it plus} the polarization, cannot go towards a liquid state.
The IMT is thus driven by the dissocation of the polarons and the screening
of the polarization potentials by the liberated electrons.  We are in a
situation which closely resembles the usual Mott transition. For finite
values of $\omega_{LO}$, we have
applied the Lindemann criterion to the PWC's melting, as was 
already done for the usual WC by Nozieres \textit{et al.} 
\cite{nozieres}. The main difference here is that one must take into account
the {\it composite
nature of the polarons}. To be able to delocalize the polarons towards a
liquid state, the quantum fluctuations for each polaron moving as a
whole, i.e.  the motion of each electron together with its surrounding
polarization, must become large in comparison with $R_s$.  On the contrary,
if the relative fluctuations of the electron with respect to the
polarization become large, the polarons break apart (dissociation).  The
Feynman treatment for the polaron \cite{feynman}, which relies on path integral
calculations, yields a natural way to evaluate such quantities: since the
polarization field in Feynman's approach is replaced by a rigid particle
with coordinate $X$ and mass $M$, one easily calculates both the quantum
fluctuations of
the center of mass $R=(m^*x+MX)/(m^*+M)$ and of the relative coordinate
$r=x-X$ ($x$ is the electron coordinate). This gives two different
Lindemann criteria, which correspond to two competing melting scenarios:
(i) ${\left < \delta R^2 \right >}^{1/2}/R_s>1/4$ for the melting 
towards a liquid state of polarons; (ii) ${\left < \delta r^2
\right>}^{1/2}/R_s>1/4$ for the polaron dissociation.

The frequencies of the two
degrees of freedom $r$ and $R$, respectively
$\omega_{int}$ and $\omega_{ext}$, were calculated in ref.\cite{quem1},
together with the ratios (i) and (ii) for different
$\alpha$.  From the basic observation that a polaron is a bound state of an
electron plus a phonon cloud, it was pointed out that the frequency of
vibration of the polaron as a whole is physically limited by the phonon
frequency ($\omega_{ext}\le \omega_{LO}$).  In other words, at strong
electron-phonon coupling (or equivalently at small $\omega_{LO}$), the lattice
polarization {\it cannot dynamically follow} the increase in kinetic energy
induced by the doping, and the quantum fluctuations are transferred to the
{\it internal} degree of freedom, breaking the polaron apart.  For that
reason, the polarons are expected to {\it dissociate} at the transition
for sufficiently high $\alpha$ ($\alpha > \alpha^*
\simeq 7.5$), rather than {\it melting} to a polaron liquid. The calculated 
critical density  for $\alpha > \alpha^*$ is  $n_c
\approx 10^{19}-10^{20} cm^{-3}$ for $m^*=1-2m_e$.

Importantly, we have also calculated the same quantities for an anisotropic
PWC ($m_c \gg m^*=m_{ab}$), as described in the introduction. 
In that case the critical coupling for polaron dissociation is 
$\alpha^* \approx 3$, and the critical density is higher than
in the isotropic case: $n_c \approx 10^{20}-10^{21}$ for $m^*=1-2m_e$.
The rather moderate values of $\alpha^*$ found in the anisotropic case 
suggest that polaron dissociation should be taken into account 
in the description of  the IMT in the cuprates.

\section{THE INSTABILITY AND ITS \\ OPTICAL SIGNATURE.}

The spectrum of the low-lying excitations (phonons) of 
the PWC can also be calculated according to the Feynman model. 
Basically, one can write down an effective 
Lagrangian for the PWC \cite{quem1,quem2}, where 
each polaron is replaced by a two-particle unit $({\bf x_i,X_i})$
localized on a Bravais lattice site ${\bf R_i}$. The Lagrangian contains
two kinds of interactions: (i) local $(x_i,X_i)$ terms (polaron
effect, plus the average local field of the remaining electrons); 
(ii) long-range dipolar interactions, both instantaneous $(x_i,x_j)$ and
retarded $(X_i,x_j)$, which  are responsible for the dispersion
in the vibrational spectrum. 

According to this semi-classical treatment, 
the phonon spectrum of the PWC is given by:
\begin{eqnarray}
\label{frequency}
\lefteqn{ \Omega({\bf k}, \lambda)  
= \frac{1}{2} \left\lbrace \phantom{\frac{1}{2}} \omega_{pol}^2 + \omega_W^2({\bf k}, \lambda)/ 
\varepsilon_\infty \right. } \\
& &\pm    
\left. \sqrt{\left[\omega_{pol}^2+ \omega^2_W({\bf k}, \lambda) \right]^2-4
\omega_{LO}^2 \omega^2_W({\bf k}, \lambda) / \varepsilon_s} \right\rbrace
\nonumber
\end{eqnarray}
where $\omega_W({{\bf k}, \lambda})$ are the eigenfrequencies
of the usual WC \cite{bagchi}, labeled by the index $\lambda$ 
(two transverse acoustical branches
and one longitudinal optical). These  satisfy the Kohn
sum rule: $\sum_{\lambda =1}^3 \omega^2_W({{\bf k}, \lambda})=
\omega_p^2$ ($\omega_p^2=4 \pi ne^2/m^*$ is the usual electron plasma
frequency, $n$ being the doping density).
The PWC has thus $6$ branches: three with the `+'
sign in (\ref{frequency}), that we call $\Omega_+$, 
all optical and which correspond to the
vibrations of the `internal' degrees of freedom (the relative motion of the
polarization with respect to the electrons); three branches with the
'-' sign, that we call $\Omega_-$, 
corresponding to the vibrations of the polarons as a whole
(electrons + polarization oscillating in phase).
Two of these branches are transverse acoustical, as in the case of the
ordinary WC. 
\begin{figure}[t]
\centerline{\resizebox{8cm}{!}{\includegraphics{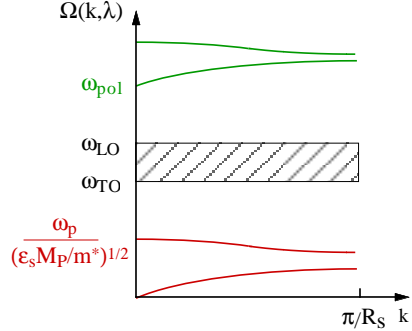}}}
\caption{The low-lying excitations of the PWC. The branches $\Omega_+$
($\Omega_-$) are located above (below) the phonons of the host lattice.}
\end{figure}
The general aspect of the  spectrum is sketched on Fig.1. In eq.
(\ref{frequency}), we have introduced an important
quantity:
\begin{equation}
\omega_{pol}^2=\left( M/m^*+1 \right) \omega_{LO}^2-
\omega_p^2/3 {\tilde \varepsilon}
\end{equation}
which corresponds to the
transverse collective modes of the `internal' degree of freedom at ${\bf k}=0$.
\begin{figure}[t]
\centerline{\resizebox{5cm}{!}{\includegraphics{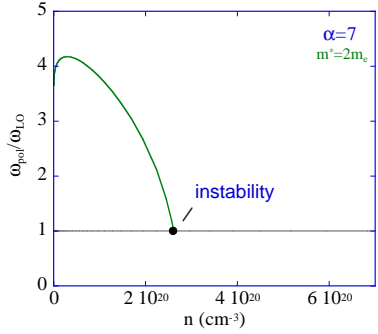}}}
\caption{The frequency $\omega_{pol}$ as a function of the
  density. The system becomes unstable when $\omega_{pol}=\omega_{LO}$.}
\end{figure}
Its evolution as a function of the density is sketched on
Fig.2. We have proved elsewhere \cite{quem2} that such excitations
can only be  stable
provided that  $\omega_{pol} > \omega_{LO}$. 
Beyond this limit, i.e. at densities higher than a certain
$n_{c1}$, the PWC is unstable with respect to  polaron dissociation.

Since  $\omega_{pol}$ is the frequency of a transverse mode at ${\bf k}=0$, it corresponds to a peak --the \textit{polaron peak} -- 
in the optical conductivity. As the density increases,
$\omega_{pol}$ decreases until $n_{c1}$ is reached,
and eventually saturates at 
$\omega_{pol} \approx \omega_{LO}$. This behaviour has indeed been 
observed in
Nd$_{2-x}$Ce$_x$CuO$_{4-y}$ \cite{lupi}.

\section{OVERSCREENING OF THE \\ COULOMB FORCE.}

The (longitudinal) dielectric constant
$\varepsilon_L \left( {\bf k}, \omega \right)$ can also be calculated.
It is given by:
\begin{eqnarray}
\label{epsilon}
\frac{1}{\varepsilon_L({\bf k}, \omega)}=
\frac {1}{\varepsilon_{ph}(\omega)} 
\left[1- \frac {\omega_p^2} {\varepsilon_{ph} (\omega)}
\times \right. \phantom{xxxxxxxxx} \\
\left. \times 
\sum_{\lambda} \frac {({\bf k}\cdot \epsilon_{{\bf k} \lambda})^2}{k^2}
\frac {\omega_{LO}^2-\omega^2}
  {[ \Omega_{-}^2({\bf k}, \lambda) -\omega^2 ] 
   [ \Omega_{+}^2({\bf k}, \lambda) -\omega^2 ]} \right] 
\nonumber
\end{eqnarray}
where $\varepsilon (\omega) = \varepsilon_{\infty} (\omega_{LO}^2-
\omega^2)/(\omega_{TO}^2- \omega^2)$, 
is the dielectric constant of the medium, 
with as usual $\omega_{LO}^2 /
\omega_{TO}^2 = \varepsilon_s/ \varepsilon_{\infty}$.  
From eq. (\ref{epsilon}), it is easy to extract the static
dielectric constant, which 
is just proportional to the static dielectric constant of the
usual WC  calculated by Bagchi \cite{bagchi}:
\begin{equation}
\label{static}
\frac {1} {\varepsilon_L ({{\bf k}, 0})} = \frac {1} {\varepsilon_s} 
\left[1- \frac {\omega_p^2}{k^2} \sum_{\lambda}
\frac {({\bf k} \cdot \varepsilon_{{\bf k} \lambda})^2} 
      {\omega_W^2({\bf k},\lambda)} \right].
\end{equation}
Owing to the Kohn sum rule, one
easily shows that $\varepsilon ({{\bf k}, 0})<0$ for any $k>0$. As it was
discussed quite generally by Dolgov {\it et al.} \cite{dolgov}, this 
does not necessarily imply an instability: it simply means that the
Coulomb repulsion between two test charges in such a system is
overscreened by the vibrations of the WC. When one approaches 
the phonon instability, the negativity
of the dielectric constant is extended on a large
region of $({\bf k}, \omega)$.
From this result, we conjecture that beyond the IMT, the moving liberated
charges can condense in a superconducting state owing to the overscreening of
their Coulomb repulsion by the remaining localized PWC. 
This can work only
up to a second critical density $n_{c2}$ at which
the localized polarons themselves are completely screened and destroyed.
In some sense, we have some kind of staggered
Mott-like transition which
begins at $n_{c1}$ and ends at $n_{c2}$. For $n<n_{c1}$ the doping
charges form polarons which crystallize. They all disappear for
$n>n_{c2}$, while for $n_{c1}<n<n_{c2}$, there could be a separation between
localized polarons and condensed charges, the glue being the long-range
Coulomb force mediated by the vibrations of the PWC.

\end{document}